\documentclass[aip,graphicx]{revtex4-1}

\usepackage{graphicx}

\usepackage[american]{babel}
\usepackage[latin1]{inputenc}

\begin{document}

\title{A Bayesian Framework for Opinion Updates}

\author{Andr\'e C. R. Martins }
\email[] {amartins@usp.br}
\affiliation{              
GRIFE - EACH, Universidade de S\~ao Paulo\\
Av. Arlindo B\'etio, 1000, S\~ao Paulo, 03828-080, Brazil
}

\date{\today}

\begin{abstract}
Opinion Dynamics lacks a theoretical basis. In this article, I propose to use a decision-theoretic framework, based on the updating of subjective probabilities, as that basis. We will see we get a basic tool for a better understanding of the interaction between the agents in Opinion Dynamics problems and for creating new models. I will review the few existing applications of Bayesian update rules to both discrete and continuous opinion problems and show that several traditional models can be obtained as special cases or approximations from these Bayesian models. The empirical basis and useful properties of the framework will be discussed and examples of how the framework can be used to describe different problems given.
\end{abstract}


\maketitle


\section{Introduction}

Opinion Dynamics~\cite{castellanoetal07,galametal82,galammoscovici91,sznajd00,stauffer03a,deffuantetal00,hegselmannkrause02} modeling lacks a clear theoretical basis. One unifying proposal exists for discrete opinion models \cite{galam05b}, but it does not include continuous opinions. A unifying framework should be able to both help us create new descriptions of new circumstances as well as, if possible, shed some light on the meaning of the already existing models. Ideally, a framework similar to that represented by Lagrangeans in Classical Mechanics, in the sense of allowing different circumstances of the world to be described within the framework, would be an important contribution to the area. Different models try to explain different aspects of the way people influence each other's opinions, often, with no common ground. While it is true that the area has observed a number of first successes in predicting a few events~\cite{bernardesetal02a,galam04}, it still lacks more testing. Models, while ingenious, are typically proposed in an ad hoc basis. 

Here, we will see that Bayesian inference can be used as a basis to develop a general framework for the updating of the opinions of agents. By clearly defining what is being discussed, how communication happens, and how likely each agent considers its neighbors to know something about the problem, Bayesian rules can be used to create models that incorporate all these features, sometimes into a simple model. We will see that the formalism is general in the sense that an update rule will always be obtained, once the relevant functions are defined, meaning that the formalism is actually quite general. And, by its dependence on the communication process, it will also allow us to better understand the meaning of traditional models.

Bayesian updating rules have been used before, both for discrete models, in the Continuous Opinions and Discrete Actions (CODA) model~\cite{martins08a,martins08b} as well as for continuous models~\cite{martins08c}. The CODA model allowed the observation of emergence of extremism, even when no extremist agents were observed initially. It will be briefly reviewed here as an example of the application of Bayesian update rules and I will demonstrate that it can be seen as a general case for discrete model, that are recovered as one specific limit of the model. For the purely continuous problem, where agents tell their full continuous opinion to each other will be studied here, Bayesian update models~\cite{martins08c} recovered the qualitative results of Bounded Confidence models~\cite{deffuantetal00,hegselmannkrause02} and allowed an extension where the threshold-equivalent variable is updated and stubbornness emerges as a consequence of the model.

The qualities of this framework will be discussed, also as a possible heuristic approximation to rational behavior. We will show that this framework can have consequences outside pure update rules, like an easier way to treat decisions and not only opinions or with a natural definition of a network in the continuous update variable problem. We will see that a variation of the CODA model can be seen as a general case of discrete update rules, when a specific limit is taken. And, finally,  the power of using Bayesian methods will be illustrated with two applications: one where the opinions lead to the breakdown of a network of trust, and another where an analysis of a case where agents share the full strength of their probabilistic opinion about a discrete choice is performed.

\section{Bayesian Update Models: The Framework}\label{sec:bayesupdate}

When people debate any issue they are interested about, the opinion of the other part can be considered as information about what choice is most likely to be best. Opinion Dynamics models are based on that assumption, as they use rules where each agent can change its opinion by the interaction with other agents. However, in general, often for the sake of simplicity, they fail to recognize the difference between internal and observed opinions. And yet, even if people were completely honest about their opinions, our language and observations certainly don't match our personal opinions. Discrete models generally opinions with no strength attached to them, while Bounded Confidence models have the agents express their correct opinions as an exact number. It is clear, however, that human language is not as precise as exact numbers \cite{urbig03}.

This distinction between internal and stated opinion can certainly have consequences for the dynamics of the system. However, at first, it might not be clear how to deal with such a distinction. While there are normative rules about how people should change their opinions in face of new information \cite{vonneumannmorgenstern47a,savage54,jaynes03,caticha04}, early laboratory experiments seem to show that people violate strongly those rules \cite{plous93,baron,allais,ellsberg61} . On the other hand, more recent analysis seem to indicate that people are not as incompetent as those experiments seem to indicate, but actually use smart heuristics to solve their everyday problems \cite{gigerenzergoldstein96,gilovichgriffin02,griffithstenenbaum06a}. In the case of probabilistic decisions, our heuristics can be understood as an approximation to a Bayesian inference \cite{martins06} and it was also observed that our inductive reasoning resemble Bayesian  reasoning closely \cite{tenenbaumetal07}. 
Therefore, it makes sense to use Bayesian updating as a basis for Opinion Dynamics models. That is so in the sense that there is some empirical evidence that, although Bayesian reasoning won't provide a perfect description of the microscopic interactions, it will give us a good approximation.

This means that one can develop Opinion Dynamics models by following simple rules, and those models can be adapted to different situations and scenarios. A simple recipe exists for any such kind of models, that is presented here as a series of steps to be followed:

\begin{enumerate}
	\item Identify what is the real debated issue that the model should represent and assign a variable $x$ to that issue. A choice between  different ideas or theories is a discrete choice and, therefore, $x$ should be, in this case, a discrete variable. If the debate is about establishing the value of a continuous variable, $x$ should be that continuous variable that the agents are talking about. Regardless of the case, the objective of each agent is to improve its inference about $x$.
	\item In order to build a Bayesian model, each agent $i$ needs to have a subjective opinion about $x$, represented by a probability distribution $f_i (x)$ that indicates agent $i$ belief on how likely each possible value of $x$ is.
	\item communication does not necessarily means stating a value for $x$. Notice that each agent has a full distribution in its mind and a complete description of his state of knowledge is given only by the distribution $f_i (x)$. Therefore, the modeler must decide how to represent the communication process. communication can be as simple as observing a choice between two alternatives by the neighbor $j$, where $j$ simply chooses the alternative with larger probability of being correct \cite{martins08a}. Or the agents can give some point estimate for $x^*$, per example,  the average $E[X]$, obtained from the distribution $f_i (x)$ \cite{martins08c}. By definition, communication means the statement of a numeric value $A_j$ by agent $j$, such that $A_j[f]$ is some functional of $f_j (x)$. In principle, agent $j$ could communicate more than one value and, in this case, $A_j$  would be a vector.
	\item The agents must have in their minds a relationship between the true value of $x$, $x^*$ and the stated value $A_j$, given by a probability distribution $p(A_j | x^*)$. That is, given that a possible value  $x^*$ is the correct value, what is the chance the neighbor $j$ will comunicate $A_j$. This model can be as simple as a fixed probability, as we will see in Section \ref{sec:coda}, or as complex as a full evaluation of the inference process of $j$ as the modeler wishes.
	\item The probability distribution $p(A_j | x)$ plays the role of a likelihood of the observation $A_j$. Since agent $i$ already had a prior opinion $f_i (x)$, obtaining its posterior opinion $f_i (x| A_j )$ is a simple task of applying Bayes Theorem \cite{jaynes03}. The posterior becomes the new opinion of $i$ and its stated values must be recalculated as $A_i[f_i (x| A_j )]$.
	\item Define who interact with whom and how often.
\end{enumerate}

In order to better understand this framework and illustrate it, two already existing Bayesian Opinion Dynamics models will be discussed bellow. They are the Continuous Opinions and Discrete Actions (CODA) model~\cite{martins08a,martins08b} and a model where communication is made by the use of a continuous variable  \cite{martins08c}, as in Bounded Confidence models. We will also see how they relate to more traditional models.

\section{Traditional existing models and Bayesian version}

\subsection{Discrete Opinions and the CODA Model}\label{sec:coda}

In the Continuous Opinions and Discrete Actions (CODA) model~\cite{martins08a,martins08b}, each agent $i$  is trying to decide between two conflicting options. That is, $x$ is a discrete variable with only two possible values, assumed here to be $\pm 1$. This means that the subjective opinion $f_i (x)$ can be trivially described as  $f_i (+1)=p_i$ and, therefore, $f_i (-1)=1-p_i$ . The communication between agents only involve stating which choice is preferred by the agent. That is, what is observed is as spin $s_i$, given by $s_i=A_i[f]=sign(p_i-0.5)$. Finally, the likelihood can be chosen in the simpler possible way, that is, each agent considers there is a chance $p(s_j=+1 | x=+1)=a>0.5$. That is, everyone assigns the same fixed chance $a$ greater than 50\% that a neighbor will choose the best alternative.

With the introduction of a social network that specifies who can be influenced by whom, the model is ready. Of course, changes of variable are often useful.
This model is much simpler when we work with the log-odds $\nu$ in favor of $+1$, defined as $\nu_i=\ln(\frac{p_i}{1-p_i})$.  Bayes Theorem causes a change in $p_i$ that translate to a simple additive process in $\nu_i$. That is, if the neighbor supports $+1$, $\nu_i$ changes to $\nu_i+\alpha$, where $\alpha=\ln(\frac{a}{1-a})$; if the neighbor supports $-1$, $\nu_i$ changes to $\nu_i-\alpha$. That is, the model is a simple additive biased random walk, with the bias dependent on the choice of the neighbors of each agent.

When the spatial structure is introduced, simulations have shown~\cite{martins08a,martins08b} that the emerging consensus is only local. Neighborhoods that support one idea will reinforce themselves and, with time, most of the agents become more and surer of their opinions, to the point they can be described as extremists. This happens even when all the agents had moderate opinions as initial conditions, unlike other models, where extremists have to be artificially introduced from the beginning. One should notice that the underlying continuous opinion allows us to speak of strength of opinions, unlike typical discrete models and, as such, at first, it is not so clear how CODA relates to those models.

When analyzed using the framework, it is clear how one can generalize CODA model to different scenarios.  Per example, by modeling a situation where $\alpha\neq\beta$ and $\beta$ is a function of time, it was possible to obtain a diffusive process from the CODA model where the diffusion slows down with time until it freezes~\cite{martinspereira08a}, with clear applications in the spread of new ideas or products. By modelling the influence of Nature as a bias in the social process of Science, CODA also proved useful to improve the understanding of how scientific knowledge might change \cite{martins10a}.

As an extension of the model, we can assume that the likelihoods depend not only on the opinion of the neighbor, but also on the agent's own observed choice. This is equivalent to introducing in the agent some awareness that its neighbor's choices might be dependent not only on the best choices, but could also be a reflection of its own influence upon that neighbor. For calculation purposes, assume, without lack of generality, that the first agent choice is $s_i=+1$. That is, the likelihood $P(s_j=+1|x=1)$ is replaced by two different probabilities 
\begin{equation}
a=P(s_j=+1|x=+1,s_i=+1)\neq P(s_j=+1|x=+1,s_i=-1)=c
\end{equation}
and $P(s_j=-1|x=-1)$ is replaced by 
\begin{equation}
b=P(s_j=-1|x=-1,s_i=-1)\neq P(s_j=-1|x=-1,s_i=+1)=d.
\end{equation}

Solving the Bayes Theorem and calculating the log-odds of the opinion, if the neighbor agrees ($s_i=+1$), we have 
\begin{equation}\label{eq:agreement}
\nu(t+1)=\nu(t)+\ln\left(\frac{a}{1-d}\right),
\end{equation}
and, if there is disagreement,
\begin{equation}\label{eq:disagreement}
\nu(t+1)=\nu(t)+\ln\left(\frac{1-a}{d}\right).
\end{equation}
The steps will only be equal in modulus, aside different signs, if $a=d$. This corresponds to the situation where both $x=+1$ and $x=-1$ are equally strong in influencing the agents and the agent $i$ choice is considered irrelevant for the choice of its neighbor $j$. On the other hand, if the agent $i$ considers that, when $s_i=+1$, it is more likely that a neighbor will choose $s_j=+1$, than we must have $a>d$. In this case, the steps will not have the same value and disagreement will have a more important impact than agreement.

The case where $a\rightarrow 1$ is interesting. If $a=1$ exactly, agent $i$ expects that, whenever it chooses $a$ and $x=+1$ is actually the best choice, the neighbor $j$ will also choose $x=+1$ with certainty. That means that an observation of $s_j=+1$ carries no new information,  while $s_j=-1$ would actually prove that $x=+1$ can not be the better choice. What happens is that, when $a=1$, the problem is no longer probabilistic, but one of Classical Logic. And as soon as the agent observes both decisions on its neighbors, it is faced with an unsolvable contradiction, unless $a$ is not exactly 1, but only close to. That is, we can work with the limit $a\rightarrow 1$, but $a$ should actually never be exactly 1.

Calculating the limits of the steps in Equations~\ref{eq:agreement} and ~\ref{eq:disagreement}, we have, for the agreement case,
\begin{equation}
\lim_{a\rightarrow 1}\left(\ln\left(\frac{a}{1-d}\right)   \right)= L,
\end{equation}
where $L$ is finite and non-zero. For disagreement, on the other hand, we have 
\begin{equation}
\lim_{a\rightarrow 1}\left(\ln\left(\frac{1-a}{d}\right)  \right)\rightarrow -\infty.
\end{equation}

That is, agreement will tend to cause a negligible change to the value of $\nu$, when compared with the change caused by disagreement. If all agents start with reasonably moderate opinions, so that, whenever they find disagreement, their choices will flip, the system, in the $a\rightarrow 1$ case is a simple one. Whenever the neighbor agrees, the first agent will not update its opinion (or update very little, if $a$ is not exactly 1). When the neighbor disagrees, the first agent will change its observed opinion to that of the neighbor. In other words, when agent $i$ observes agent $j$ choice, it always end with the same choice as $j$. In the limit, we obtain the traditional voter model~\cite{cliffordsudbury73,holleyliggett75}. 

That is, we have a dynamics where the agent only updates its mind when there is disagreement. This same update dynamics is observed in other discrete models, as per example, for Sznajd interactions~\cite{sznajd00,stauffer03a,sznajd05}. In Sznajd model, it takes two agreeing agents to convince all other neighbors. Basically, it works the same way as the voter model, except for the description of when an interaction happens. Since the Bayesian framework is only applied here to the opinion update and not to the rules of interaction, we have the same case as we had in the voter model. Other features, such as contrarians~\cite{galam04}, are also easily introduced by a simple change in the likelihood. If an agent considers its neighbor more likely to be wrong than correct, the agent opinion will change away from that of the neighbor, hence, a contrarian \cite{martinskuba09a}. 

Finally, the models of hierarchical voting~\cite{galam2003a,galam06b}, where the decision of each level is obtained from the majority of the voters, except when there is a tie, can also be easily translated into CODA Bayesian language using the same strategy as in the voter model. That is, agreement with the majority means no reinforcing of previous opinion, while disagreement leads to an observable change. If there is a slightly different likelihood in favor of one theory, when there is a tie, that theory will tend to be picked up. The same effect could also happen due to small differences in the probabilistic continuous views of the individuals in the tie groups. Interestingly, the translation of the problem into CODA formalism suggests natural extensions of the model, where the final opinion might depend also on the continuous probability each agent assigns to each proposition.

\subsection{Continuous Variables}

We can also investigate the relation between the framework and the Bounded Confidence continuous opinion models\cite{deffuantetal00, hegselmannkrause02}. That relation was studied in a previous work
\cite{martins08c} and, for the sake of completude, it is reviewed here, as that Bayesian model will be extended in the next Section. 

In this case, the agents want to learn the value of a continuous variable $\theta$, that plays the part of $x$ in the framework description. Each agent $i$ has a continuous prior opinion about a variable $\theta$, represented by a prior distribution $f_i(\theta)$, with an average estimate of  $\theta$ given by $x_i=E[\Theta]$. Here the average $x_i$ is the value that is communicated to the other agents. Assume also the prior is a Normal distribution with uncertainty $\sigma_i$. As the idea was to get a model comparable to Bounded Confidence models, it is necessary to introduce a likelihood that is a mixture of the Normal distribution (probability $p$), with the same uncertainty the agent assigns to his own prior, and a Uniform distribution over the whole range of possible values.
\begin{equation}\label{eq:likelihooddecept}
f(x_j|\theta) = p N(\theta,\sigma_{j}^{2}) + (1-p) U(0,1),
\end{equation}

 That is, the agent thinks that the other agent might know something (Normal around the true value, with probability $p$) or know nothing, just stating a completely random guess (Uniform, $1-p$). Using these rules the new average, after interacting with agent $j$ will be

\begin{equation}\label{eq:averagedecept}
x_i(t+1)=p^*\frac{x_i(t)+x_j(t)}{2}+(1-p^*)x_i(t)
\end{equation}
where
\begin{equation}\label{eq:posteriorp}
p^* = \frac{p\frac{1}{\sqrt{2\pi}\sigma_i} e^{-\frac{(x_i(t)-x_j(t))^2}{2\sigma_{i}^{2}}} }{p\frac{1}{\sqrt{2\pi}\sigma_i} e^{-\frac{(x_i(t)-x_j(t))^2}{2\sigma_{i}^{2}}} +(1-p)}.
\end{equation}

Assuming only the average estimate $x_i$ is updated, instead of the whole distribution, this generates a model with all the qualitative features of Bounded Confidence models. If one approximates Equation~\ref{eq:posteriorp} by a step function, we actually do recover Bounded Confidence exactly.

Of course, better updatings, from the point of view of rationality, are possible. Ideally, each agent should update the whole probability distribution, but one natural next step is simply to add the updating of the second moment of the distribution (that way, one can still have Normal distributions at every step). We have then
 
 \begin{equation}\label{eq:varupdate}
\sigma_{i}^{2}(t+1)=\sigma_{i}^{2}(t)\left( 1-\frac{p^*}{2} \right) + p^*(1-p^*) \left(\frac{x_i(t)-x_j(t)}{2}\right)^2.
\end{equation}

Now the agents have a dynamics for their uncertainty and it is observed that they become more stubborn with time. This extra updating allows us to notice another interesting property of Bayesian rules. Since each parameter has a clear interpretation, it is easier to propose natural extensions of the model. We will return to this model and propose a simple new extension to better illustrate this in Section \ref{sec:networks}.

\subsection{Rationality and Bayesian rules}

A comment on the possibility of modeling rational agents is needed here. It is true that Bayesian update rules provide a way to model how agents change their minds. However, it is important to notice that the agents described here are not completely rational. Full rationality would imply much more than just using Bayes Theorem to update subjective probabilities. One of the most simple and important characteristics one expects from full theoretical rationality is that a rational individual should analyze all information available as well as possible. And that is not done by the agents in any of the models discussed here. Agents in the models presented here use heuristics, just as humans are believed to do~\cite{gilovichgriffin02}.

Alse, some values were assumed as known without uncertainty, as the chance $a$ that a neighbor will favor the best hypothesis. In a real problem, a rational agent must also solve the problem of testing different likelihoods. A rational agent should also model the way the neighbors makes up their minds. That would include how each neighbor is influenced by its own neighbors, including the original agent. Such model would require a model about how the neighbor supposes the first agent decides the best option and, if the neighbor were also rational, that would include the agent having to model how the neighbor models the opinions of the agent. Approximations to full rationality are needed. But they are needed for humans to reason also, as our cognitive ability is not infinite. And simple Bayesian updating rules can suggest new rules as well as alterations in the models, if one deems them necessary. They also help explain known results better, as we have seen in Section~\ref{sec:bayesupdate}.

\section{Using the Framework}

In other to explore the power of the proposed framework better, two toy models will be discussed bellow that are alterations of the continuous model and of the CODA model.

\section{Examples}

\subsection{Networks}\label{sec:networks}

Equation~\ref{eq:posteriorp} is, as written and used in the continuous opinion model, just a step in calculating the new average. However, it refers to how trustworthy the agents think the other agents are. If agents not only change their minds about the variable they want to learn about, but also keep the information about the trustworthiness of each agent they have interacted with, we will have a $p_{ij}$ that evolves with time and that represents what each agent $i$ thinks of the others. This is, a network of trust will evolve. For this model, simulations show that updating that value introduces no significant quantitative differences in the outcome of the model when only final opinion is analysed. We see that the trust network splits into the same number of values of final opinions, for a given choice of $\sigma_i$, with $p_{ij}=0$ trust between agents with different final values of $x$ and with $p_{ij}=1$ for agents with the same opinion. The reason that the updating of $p_{ij}$ has negligible influence in the final opinion distribution is that $p_{ij}$ has a much weaker in the update in Equation~\ref{eq:posteriorp} than the effect one obtains from different values of $\sigma_i$.

\begin{figure}[bbbh]
 \includegraphics[width=0.45\textwidth]{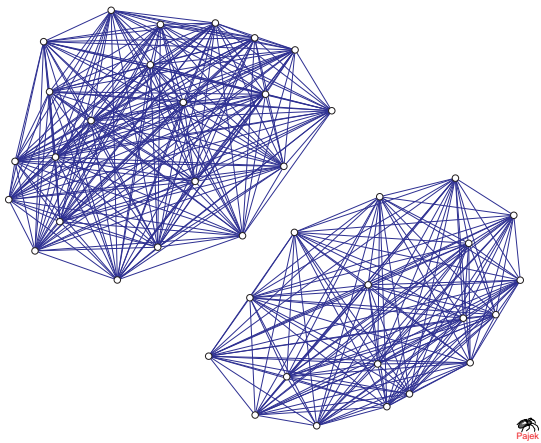}
 \caption{Network of trust after final split of opinions for $N=40$ agents. The initial uncertainty of every agent was $\sigma_i =0.06$}
 \label{fig:network}
 \end{figure}

This updated trust by use of Equation~\ref{eq:posteriorp} can be a basis for the evolution of the network of trust. If each agent has an individual opinion about the probability each one of the other agents knows what they are talking about, $p$ must be replaced by the field opinion $p_{ij}$ agent $i$ has about the correctness of agent $j$. By keeping track of these values and updating them at each interaction, Equation~\ref{eq:posteriorp} provides the dynamics for $p_{ij}$. Since $p_{ij}$ measures the trust between the agents, it is natural to understand it as a natural measure for a network of trust in the system. 

Figure~\ref{fig:network} shows the final network for the case where $N=40$ agents were left to interact, from an initial uncertainty of $\sigma_i =0.06$. Visualization was performed with the use of the Pajek software~\cite{nooyetal05a}, using the Fruchterman--Reingold energy routine~\cite{fruchtermanreingold91a}. The spread of the opinions in the simulations was basically the same as that previously obtained with no update in $p_{ij}$. Basically, $p_{ij}$ introduces a linear distrust, while the effects from a small value of $\sigma_i$ are felt as the tails of a Normal distribution and, therefore, far stronger. That is consistent with previous analysis of the model~\cite{martins08c}. What is interesting to notice is the complete split of the network into two unconnected parts, corresponding to the two different final opinions. Within each unconnected portion, agents trust each other with probability $p=$, and they all distrust those with different choices, so that $p=0$ in those cases. We can also see how the fact that we understand the meaning of the variables in the Bayesian framework has allowed us to propose a natural extension of the model.

\subsection{Discrete Choices with Continuous Verbalization}~\label{sec:concoda}

Finally, in order to better explore the formalism and also in order to make some of its properties clearer, a variant of the CODA model will be discussed where the communication between the agents is not a discrete spin value, but the full probability $p_i$ agent $i$ assigns to the possibility that the right choice is $x=+1$. Notice that the fact that the communication is continuous does no imply that $x$ should be. We still have a problem with only two possible choices $x=\pm 1$. However, the continuous probabilistic value is the communicated information. This distinction is a very important albeit neglected one. In continuous models, it is usually assumed that both the communication and the decision are continuous, but that doesn't have to be the case.

As the communication phase in the framework was changed, we need now a new likelihood, that neighbor agent $j$ will issue the value $p_j$ in the case where $x=-1$ and in the case where $x=+1$, that is, functions $f(p_j|A)$ and $f(p_j|B)$. Since all values for $p_i$ are limited to $0\leq p_i \leq 1$, the simplest choice is to take Beta distributions $Be(p_j|\alpha,\beta)$ as priors.

\[
Be(p_j|\alpha,\beta)=\frac{1}{B(\alpha,\beta)}p_{j}^{\alpha-1}(1-p_{j})^{\beta-1}
\]
where $B(\alpha,\beta)$ is obtained from Gamma functions by
\[
B(\alpha,\beta)=\frac{\Gamma(\alpha)\Gamma(\beta)}{\Gamma(\alpha+\beta))}.
\]

The Beta distribution is the prior conjugate distribution to a Binomial likelihood. That means that $\alpha$ is associated with the observed number of successes in a Binomial trial and $\beta$ with the number of failures. This translates in the fact that, if an agent thinks  that
 $p_i>0.5$, we must have $\alpha_i>\beta_i$,  as well as $\alpha_i<\beta_i$  $p_i<0.5$.

As Beta function is symmetric in $\alpha$ e $\beta$, if we want to keep the symmetry between $x=+1$ and $x=-1$, we must have for the likelihoods that, if $f(p_j|x=1)=B(\alpha,\beta)$, then $f(p_j|x=-1)=B(\beta,\alpha)$. It is also reasonable to assume that, under uncertainty conditions, it is likely that the neighbors will get a wrong answer, that is, the likelihoods shouldn't be too different. This can be achieved by making $alpha$ and $beta$ not too different, per example, $\alpha=\beta+1$. By applying the Bayes Theorem to this problem, agent $i$, when observing $p_j$,  will update $p_i$ to $p_i(t+1)$, 
\begin{equation}\label{eq:continuouscoda}
	p_i(t+1)=\frac{p_i p_j}{p_i p_j + (1-p_i)(1-p_j)}.
\end{equation}
The normalization factors of each term cancel out and Equation \ref{eq:continuouscoda} is exact. Furthermore, if we adopt the same transformation of variables as in CODA model and estimate $p_i/(1-p_i)$ we will see that the denominators also cancel and we have that
\begin{equation}\label{eq:logoddcontinuouscoda}
\ln (\frac{p_i(t+1)}{1-p_i(t+1)}) =\ln (\frac{p_i(t)}{1-p_i(t)})+\ln (\frac{p_j(t)}{1-p_j(t)})
\end{equation}

Defining  $\nu_i=\ln (p/(1-p))$, Equation \ref{eq:logoddcontinuouscoda} can be rewritten in a more elegant fashion
\begin{equation}\label{eq:logoddcontinuouscodanu}
\nu_i(t+1)=\nu_i(t)+\nu_j
\end{equation}
This is similar to the CODA dynamics, except that now, at each step, instead of adding a term that is constant in size and only varies in sign, we add exactly the log-oods of the opinion of the neighbor.

\begin{figure}[ht]
 \includegraphics[width=0.8\textwidth]{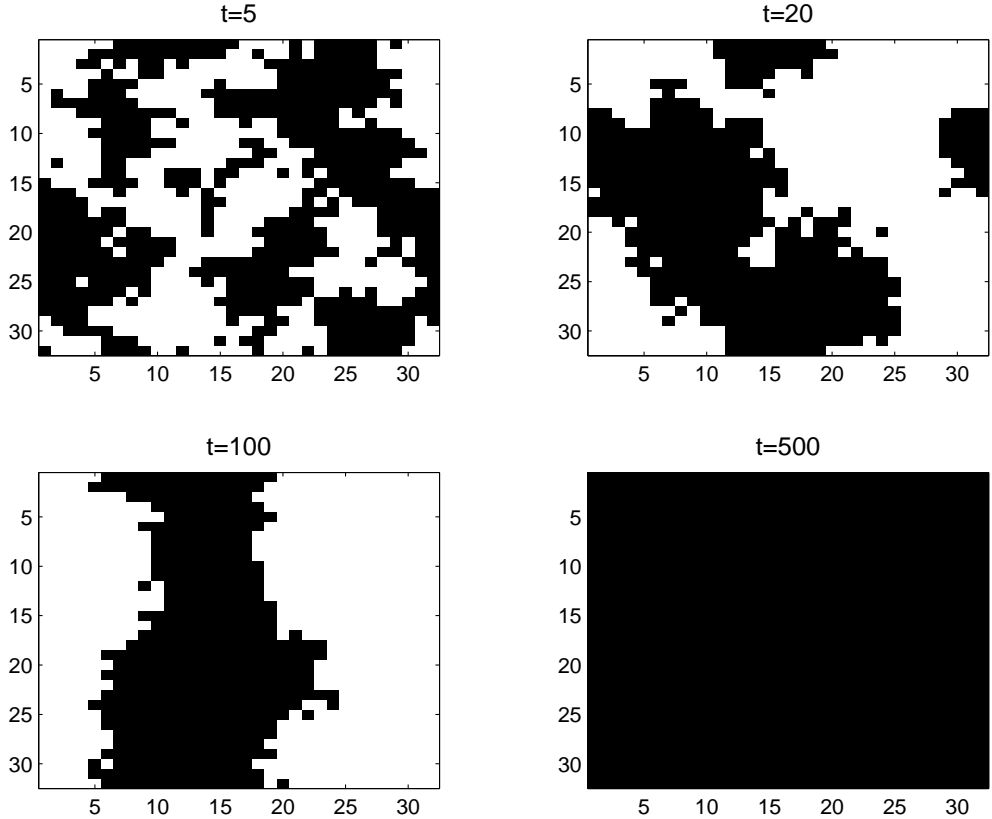}
 \caption{Typical lattice configuration showing the choices after $t$ interactions per agent.}\label{fig:contcodanlat32finalconf}
 \end{figure}

Equation ~\ref{eq:continuouscoda} allows us to search for solutions that are fixed points, what happens when $p_i(t+1)=p_i(t)$.By replacing this condition in Equation \ref{eq:continuouscoda} we have the stable fixed solutions $p_i=0$ ou $p_i=1$ for every $i$. There is also another trivial solution  $p_j=0.5$ for every $i$, but this solution is not stable. This seems to indicate that the system will tend to the extreme values, in opposition to the models where both verbalization and decision were continuous. These are the same stable points we had in CODA, but in the original model, the system was prevented from ever reaching them, except locally.

\begin{figure}[ht]
 \includegraphics[width=0.8\textwidth]{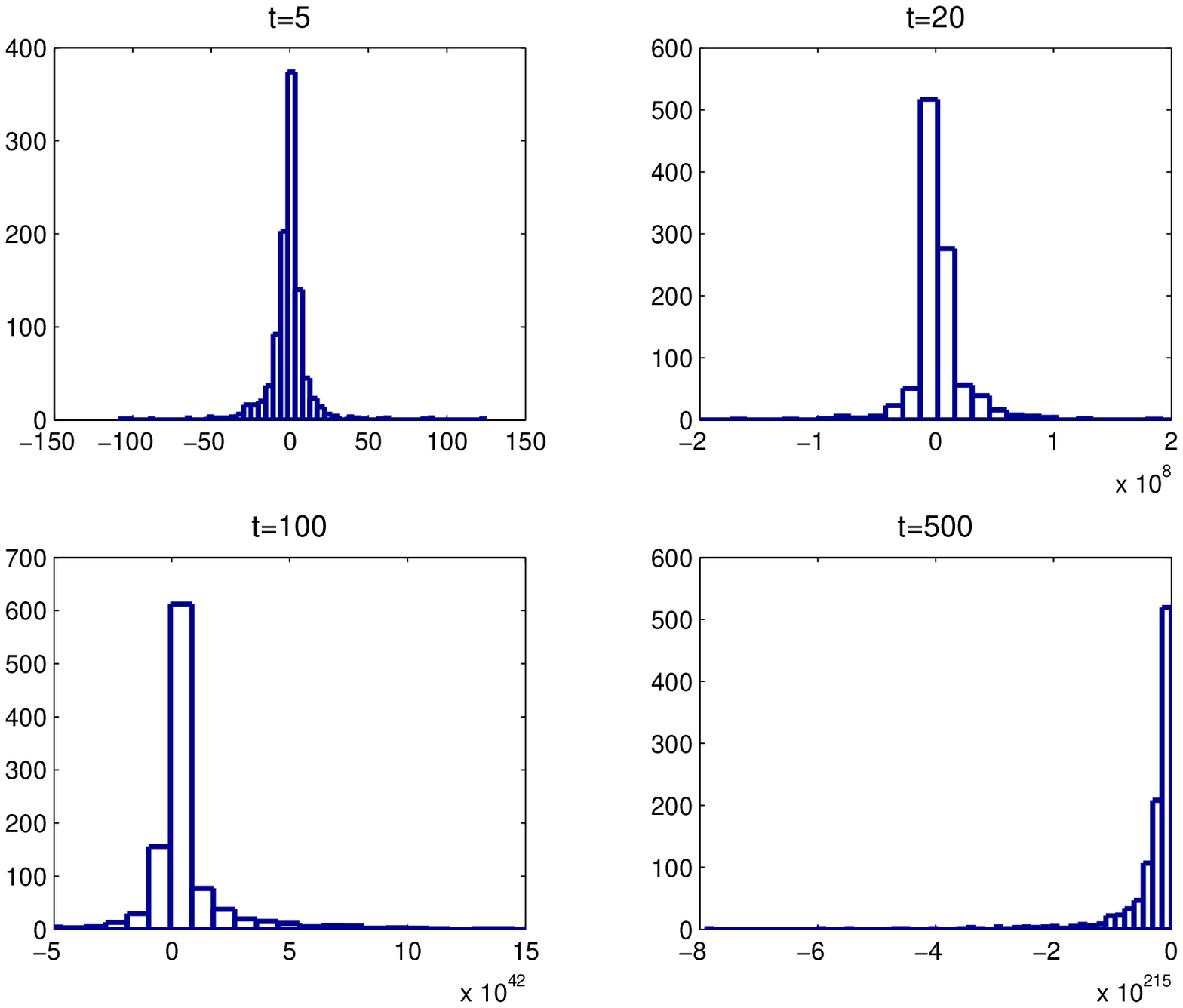}
 \caption{Distribution of the logodd opinions $\nu_i$ after $t$ interactions per agent.}\label{fig:contcodanlat32distr}
 \end{figure}

Simulations were prepared to confirm this prediction. Square lattices with periodic boundary conditions and von Neumann neighborhood were used and the state of the system observed after different average number of interactions $t$ per agent. The results for the evolution of the configurations of choices can be seen in 
Figure~\ref{fig:contcodanlat32finalconf}. At first, we observe a behavior very similar to that of the CODA model, with a clear appearance of domains with different choices. However, instead of freezing, those domains keep changing and expanding and, eventually, one of the options emerge as victorious and the system arrives at a consensus.

\begin{figure}[ht]
 \includegraphics[width=0.8\textwidth]{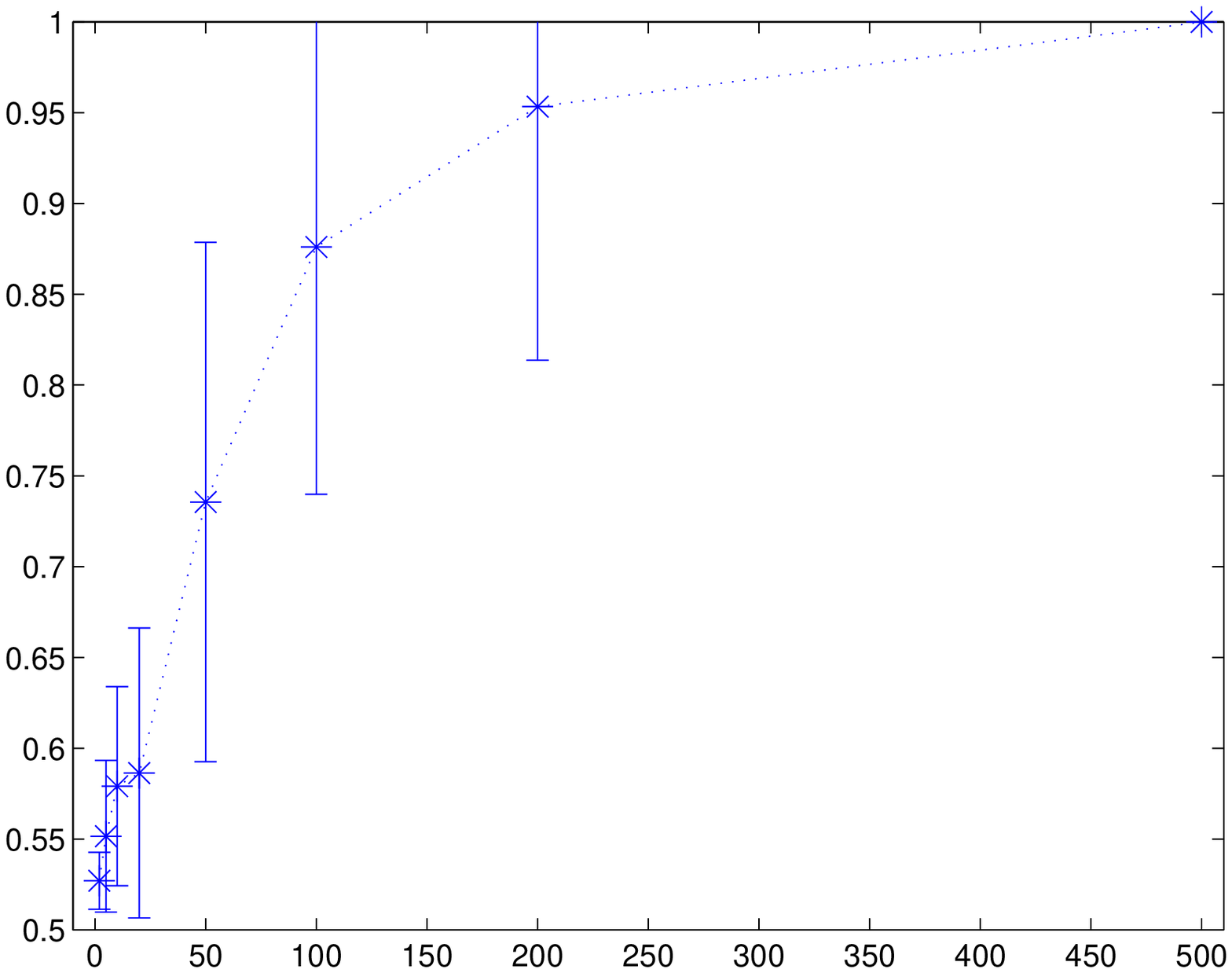}
 \caption{Average proportion of agents that agree with the majority as a function of time  $t$, measured as the average number of interactions per agents. A square lattice with  32x32 agents  was used here.}\label{fig:contcodanlat32finalprop}
 \end{figure}

Figure \ref{fig:contcodanlat32distr} shows histograms with the observed distributions for  $\nu_i$, equivalent to each of the cases in  Figure \ref{fig:contcodanlat32finalconf}. The important thing to notice there is how the scale of typical values for $\nu_i$ changes with time. In the original CODA model, the extremist peaks corresponded to  $\nu_i$  around $-430$ (that is, 500 steps with the size of the step determined by $a=70\%$). That value was already quite extreme since it corresponded to probabilities of around $10^-{300}$. 
Meanwhile the values of $10^{215}$ observed at Figure \ref{fig:contcodanlat32distr} are $\nu_i$ values, and not probability $p_i$ values. That means that opinions are incredibly extreme. The system as a whole observes a surprisingly fast appearance of extremist. After as little as 5 average interactions per agents, we have opinions so extreme that would take 100 CODA model interactions. 

What is happening here is that, as $\nu_i$ gets large by the same local reinforcement process as in CODA,  this causes larger values of $\nu_j$ to be added in Equation \ref{eq:logoddcontinuouscodanu}. And those ever increasing extreme opinions eventually reach the borders and there, they can have a very strong effect on the opinions of the agents who have taken the opposite choice. Instead of the local opinions tendencies be determined by the local configuration, in this model, an extreme opinion can actually have a much stronger impact than any others. This causes the domain walls to shift and, after some time, consensus emerges. However, this is a consensus associated with a not warranted level of certainty. 
Figure \ref{fig:contcodanlat32finalprop} shows the average proportion of agents that agree with the majority as a function of time  $t$, measured as the average number of interactions per agents. Each point shows the average result of 20 realizations and the error bars show the standard deviation of the observed values.  The tendency towards consensus as time goes by becomes obvious and, after $t=500$, all simulations ended in full consensus.

When compared with CODA model, we can conclude that it was not the fact that agents only observed discrete choices that caused the extreme opinions. Observing the strength of those opinions can have an even stronger effect into the appearance of extreme points of view. This extremism seems to be associated actually with the problem that the choice is discrete, between two competing points of view, with no compromise possible.

\section{Conclusions}

We saw that Bayesian rules can provide a theoretical basis to model the change in the opinion of agents in both a more realistic and more flexible way, probably a little closer to how real people think. The framework does not tell us how people will interact, what the variables are and how they communicate, but it makes clear that all those are important parts of the process and allows them to be introduced in any new model. It is up to the modeler to determine how any system of interest really works and, therefore, the functions and details of how the framework will be applied. In that sense, it works in a similar way to Lagrangeans in Mechanics. Which Lagrangean function must be used to describe which system is a question that must be answered by the scientists working in the field, not by the formalism. The same happens here.

The introduction of these ideas in a Ising-like scenario, where only binary choices can be observed, had made it possible the modeling of the emergence of extremism in a previous work, in the context of the CODA model. Here, we have also seen how several traditional models of the literature can be obtained as a limit case of the CODA model. 
By applying the same framework to models of continuous opinions,  it is possible to understand the Bounded Confidence models as an approximation to a Bayesian update rule when the agents keep their uncertainty unchanged. The model developed from those rules is similar, but not identical, to the Gaussian Bounded Confidence Model~\cite{deffuant06}. 

It is important to stress once more that the agents in the models are not really Bayesian agents, so no claim that humans are exactly Bayesians is implied. Once the approximate interaction rules are found, the agents in the framework follow them in a dumb way. Modeling agents with a less limited rationality can be an interesting project and it would easier to implement with Bayesian rules, but that was not the approach of this paper. On that line, it should be particularly interesting to investigate extensions of the model using perceptrons, per example, with the use of Hebbian algorithms\cite{vicenteetal98a} and work in that line is currently being conducted~\cite{vicenteetal08b}. 

What we have is that the Bayesian framework provides a good approach to creating opinion models. It can help creating better, but still simple, interaction rules to be used when studying social problems. As a framework, it is both a generalization of several traditional models and something that can be used once first principles of human interactions become clear. Another interesting aspect is that natural extensions of the model are suggested by the framework. We have explored extensions as the network of trust and the continuous communication in discrete choice problems. Other extensions are certainly possible, simply by making different approximations to a rational decision. And not only opinions can be modeled that way, but also choices and actions, thus expanding the reach of Opinion Dynamics.

\section{Acknowledgments}

The author would like to thank both Funda\c{c}\~ao de Amparo \`a  Pesquisa do Estado de S\~aoPaulo (FAPESP),  under grant 2008/00383-9, And Conselho Nacional de Pesquisa (CNPq), under grant 475389/2008-5, for the support to this work.

\bibliography{biblio}

\end{document}